\documentclass[iop]{emulateapj}
\usepackage{subfigure}
\makeatletter
\def\aoexp{18}
\def\keckstarmatches{68}

\def\eightfourteenonesixtyrarms{0.023$''$}
\def\eightfourteenonesixtydecrms{0.042$''$}

\def\keckrarms{0.057$''$}
\def\keckdecrms{0.055$''$}
\def\sncoordsme{\mbox{$\alpha$ = 9$^{\rm h}$55$^{\rm m}$42.137(9)$^{\rm s}$}, \mbox{$\delta$ = +69$^{\circ}$40$'$25.40(5)$''$} (J2000.0)}

\def\aodim{\mbox{$16'' \times 16''$}}

\def\zhengdiscov{January 14.75 ($\pm$ 0.21 days)}
\def\goobarebv{$E(B-V)_{\rm SN} = 1.22 \pm 0.05$ mag}
\def\goobarRv{$R_V = 1.40 \pm 0.15$}
\def\goobarext{($R_V^{\rm M82} = 1.4$; $A_V^{\rm M82} = 1.7$ mag)}
\def\pARv{$R_V^{\rm M82} = 3.1$ mag}
\def\pAAv{$A_V^{\rm M82} = 1.7$ mag}
\def\pBRv{$R_V^{\rm M82} = 3.1$ mag}
\def\pBAv{$A_V^{\rm M82} = 1.7$ mag}
\def\pCRv{$R_V^{\rm M82} = 2$ mag}
\def\pCAv{$A_V^{\rm M82} = 2.5$ mag}
\def\distmod{$27.73 \pm 0.02$ mag \citep{jacobsrizzi09}}

\def\aoutdate{25 January 2014}
\def\fosseyutdate{21 January 2014}
\def\MWebv{$E(B-V)_{\rm MW} = 0.14$ mag}
\def\offsetusatel{0.08$''$}
\def\ebvrsoph{$E(B-V)=0.73$}

\submitted{Submitted to The Astrophysical Journal}

\begin{document}

\title{Constraints on the Progenitor System of the Type Ia Supernova 2014J from Pre-Explosion {\it Hubble Space Telescope} Imaging}

\shorttitle{}

\email{pkelly@astro.berkeley.edu}

\author{Patrick L. Kelly\altaffilmark{1}}
\author{Ori D. Fox\altaffilmark{1}}
\author{Alexei V. Filippenko\altaffilmark{1}}
\author{S. Bradley Cenko\altaffilmark{2}}
\author{Lisa Prato\altaffilmark{3}}
\author{Gail Schaefer\altaffilmark{4}}
\author{Ken J. Shen\altaffilmark{1,5}}
\author{WeiKang Zheng\altaffilmark{1}}
\author{Melissa L. Graham\altaffilmark{1}}
\author{Brad E. Tucker\altaffilmark{1,6}}

\altaffiltext{1}{Department of Astronomy, University of California, Berkeley, CA 94720-3411, USA}
\altaffiltext{2}{NASA/Goddard Space Flight Center, Code 662, Greenbelt, MD 20771, USA}
\altaffiltext{3}{Lowell Observatory, 1400 West Mars Hill Road, Flagstaff, AZ 86001, USA}
\altaffiltext{4}{The CHARA Array of Georgia State University, Mount Wilson Observatory, Mount Wilson, CA, 91023, USA}
\altaffiltext{5}{Einstein Fellow}
\altaffiltext{6}{Research School of Astronomy and Astrophysics, The Australian National University, Weston Creek, ACT 2611, Australia}


\begin{abstract}

We constrain the properties of the progenitor system of the highly reddened Type Ia supernova (SN) 2014J in Messier 82 (M82; $d \approx 3.5$~Mpc). We determine the SN location using Keck-II {\it K}-band adaptive optics images, and we find no evidence for flux from a progenitor system in pre-explosion near-ultraviolet through near-infrared {\it Hubble Space Telescope (HST)} images. Our upper limits exclude systems having a bright red giant companion, including symbiotic novae with luminosities comparable to that of RS~Ophiuchi.   While the flux constraints are also inconsistent with predictions for comparatively cool He-donor systems ($T \lesssim$ 35,000~K), we cannot preclude a system similar to V445 Puppis. The progenitor constraints are robust across a wide range of $R_V$ and $A_V$ values, but significantly greater values than those inferred from the SN light curve and spectrum would yield proportionally brighter luminosity limits. The comparatively faint flux expected from a binary progenitor system consisting of white dwarf stars would not have been detected in the pre-explosion {\it HST} imaging. Infrared {\it HST} exposures yield more stringent constraints on the luminosities of very cool ($T < 3000$~K) companion stars than was possible in the case of SN~Ia~2011fe. 

\end{abstract}

\maketitle 

\section{Introduction}

The exceptional luminosity of Type Ia supernovae (SN~Ia), and the tight empirical relationships among the decline rate, color, and peak luminosity of their light curves \citep{ph93,ri96}, make SN~Ia useful probes of the cosmic expansion history \citep{riess98,perlmutter99}.
SN~Ia spectra and inferred $^{56}$Ni masses (e.g., \citealt{mazzali07})  
show reasonable agreement with models of the thermonuclear explosions of carbon-oxygen white dwarfs (\citealt{hill00}; \citealt{kas05}; \citealt{kas07}; \citealt{kas09}).
Additional evidence for a comparatively old progenitor population comes from the presence of SN~Ia in 
passive galaxies, and the observation that they show no preference for the brightest regions of their hosts,
in contrast to core-collapse explosions that also exhibit H- and He-deficient spectra (\citealt{kel08}; see also \citealt{ras09}).

For sufficiently nearby SN~Ia ($d \lesssim 10$~Mpc), pre-explosion {\it HST} imaging has the 
sensitivity to detect several classes of candidate progenitor systems.
Current constraints suggest that SN~Ia progenitor systems consist primarily of either binary white dwarfs \citep{ibe84,web84,shenbildsten14}, 
or binaries where a single white dwarf accretes matter from a stellar companion (\citealt{whe73}; \citealt{hanpodsiadlowski04}).
For the latter, single-degenerate channel, the white dwarf gains matter from a companion star
up to a point where its mass is close to the Chandrasekhar limit (1.4~M$_{\odot}$), precipitating eventual thermonuclear runaway. 
Accretion onto a white dwarf primary can occur through Roche-lobe overflow (RLOF)
from a secondary with a H envelope \citep{vandenheuvelbhattacharya92} or from a He star \citep{nomoto82,yoonlanger03,wangmeng09,liuchen10,geiermarsh13}.
Alternatively, in the case of the symbiotic channel, the white dwarf accretes mass from the wind generated by the secondary \citep{munarirenzini92,patatchugai11}. Characterizing the diversity of SN~Ia progenitor systems 
may be useful for explaining evidence that the luminosities of SN~Ia have a $\sim0.1$ mag dependence on the properties of the host galaxy, after correcting for light-curve shape and color \citep{kel10, sullivan10, lampeitl10, childressaldering13}. 

While earlier analyses have found useful nondetections at SN~Ia explosion sites (e.g., \citealt{maozmannucci08}; \citealt{nelemansvoss08}), \citet{libloom11} were able to place significantly fainter limits on the luminosity at the explosion site of SN~Ia~2011fe in M101 ($d \approx 6.4$~Mpc; \citealt{shappeestanek11}).
Their nondetection excludes a bright red giant companion, specifically both model Galactic progenitor symbiotic systems RS~Ophiuchi (RS~Oph) and T~Coronae Borealis (T~CrB), as well as
the He-star system V445 Puppis (V445~Pup).

Here we report constraints on the progenitor system of SN~2014J using pre-explosion near-ultraviolet (UV) through near-infrared (IR) {\it HST} 
imaging of the explosion site whose coordinates we measure using Keck-II adaptive optics (AO) imaging. Section~\ref{sec:discovan} provides a brief summary of the discovery and early analysis of
the spectra and light curve of SN~2014J. 
In \S \ref{sec:data}, we describe the Keck AO and {\it HST} pre-explosion images
that we analyze in this paper. 
The methods we use to extract upper luminosity limits for the progenitor system are explained in \S \ref{sec:methods},
while \S \ref{sec:results} presents constraints on possible progenitor systems.
Section~\ref{sec:conclusions} provides a summary of our conclusions.

\begin{figure*}[t]
\centering
\subfigure{\includegraphics[angle=0,width=3.3in]{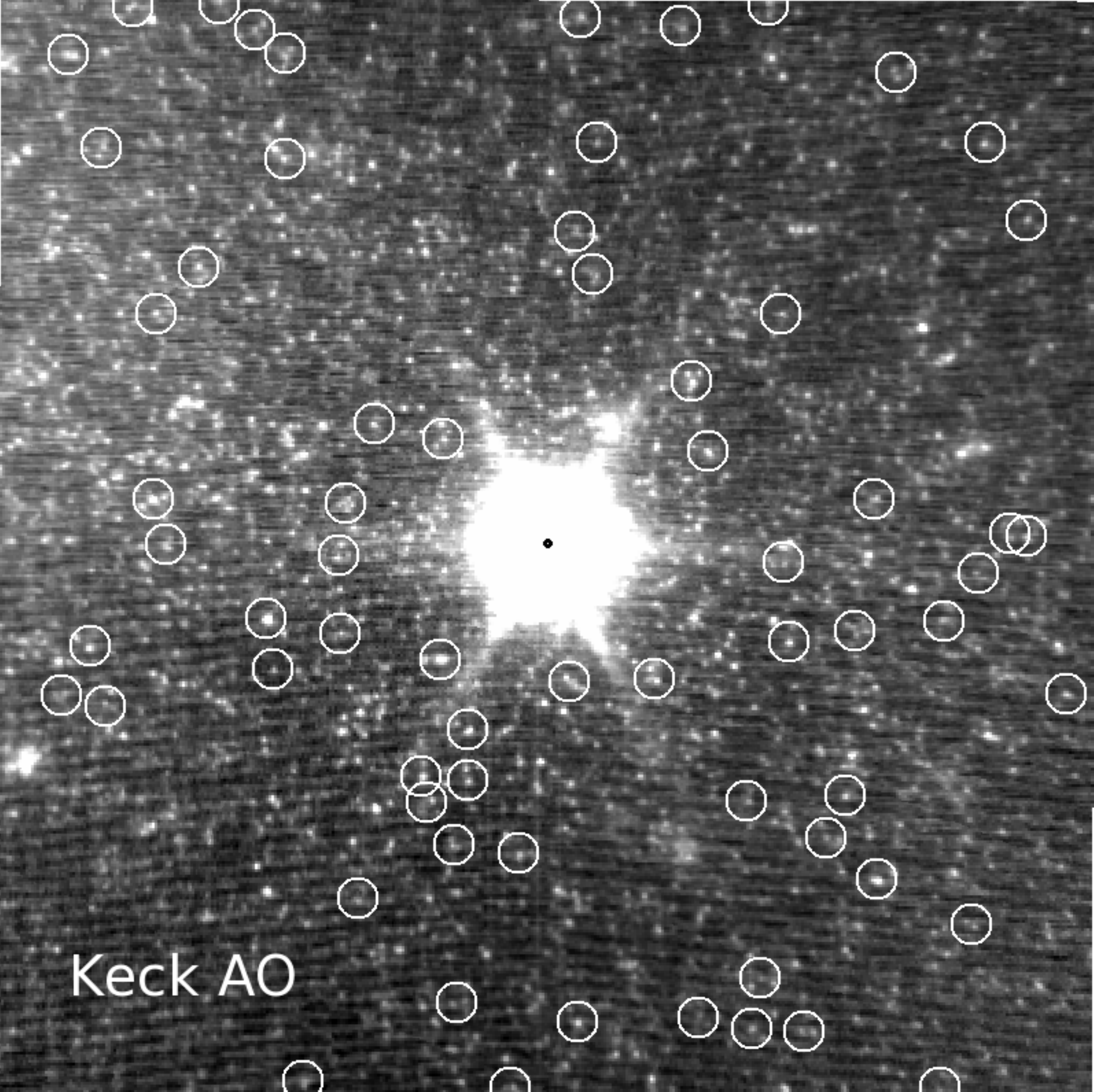}}
\subfigure{\includegraphics[angle=0,width=3.3in]{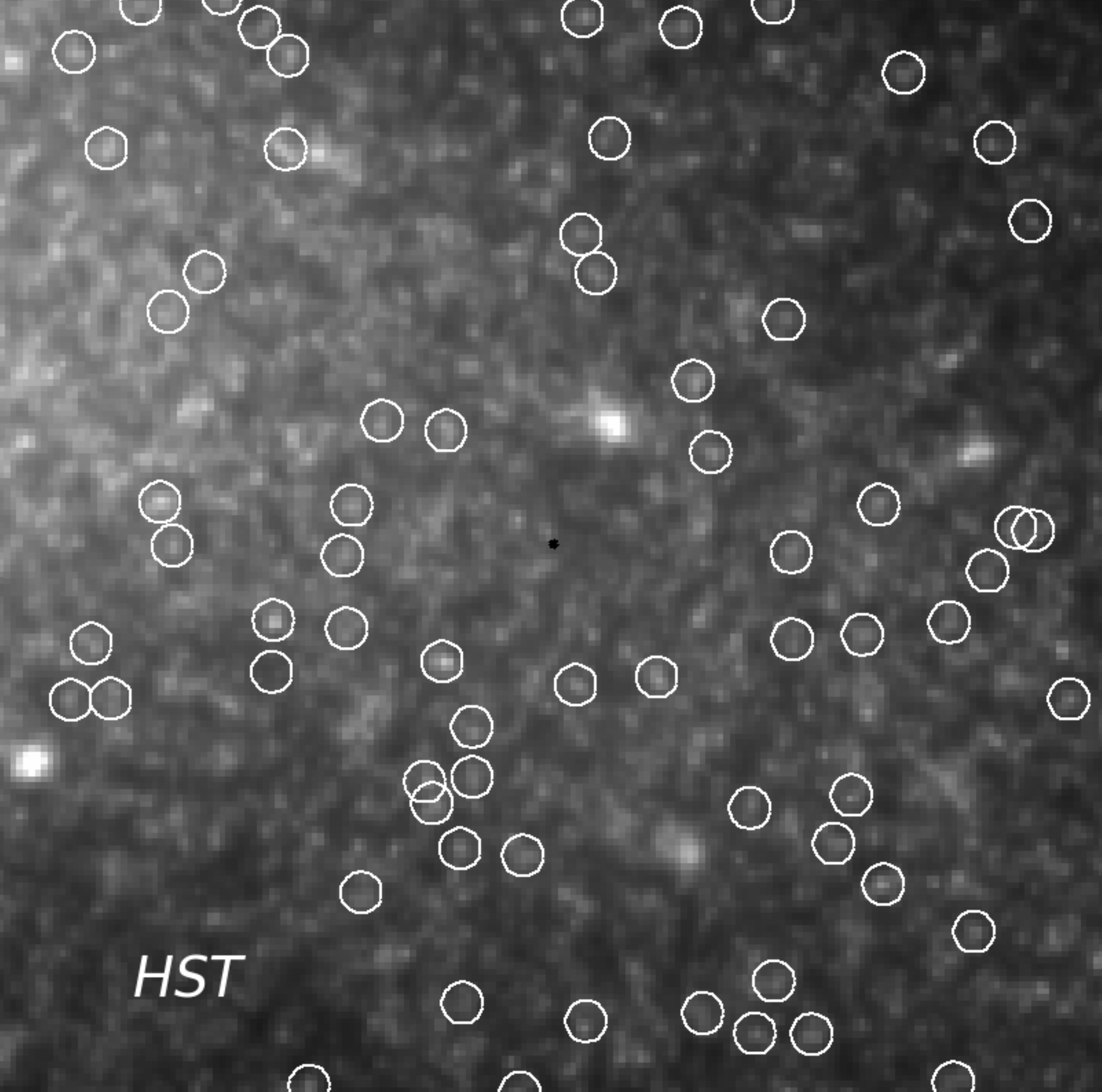}}
\caption{Coadded Keck-II {\it K}-band NIRC2 AO (left) and {\it HST} pre-explosion F160W (right) exposures of the location of SN~2014J. We use only the central \aodim~of the distortion-corrected AO image to perform astrometric 
registration. The \keckstarmatches~sources used for registration are identified with white circles, while the 
position of SN~2014J is marked by a black circle with radius corresponding to the uncertainty in that position estimate. }
\label{fig:AO}
\end{figure*} 

\begin{figure*}[t]
\centering
\subfigure{\includegraphics[angle=0,width=3.25in]{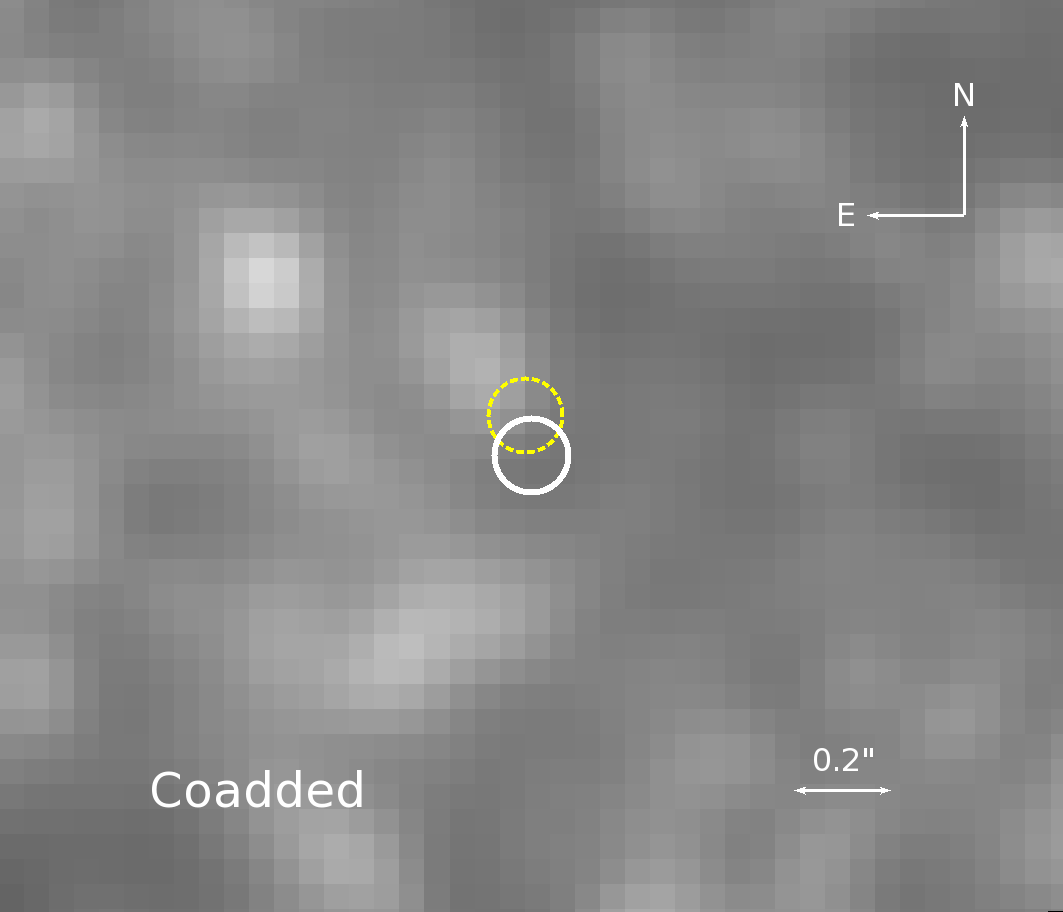}}
\subfigure{\includegraphics[angle=0,width=3.25in]{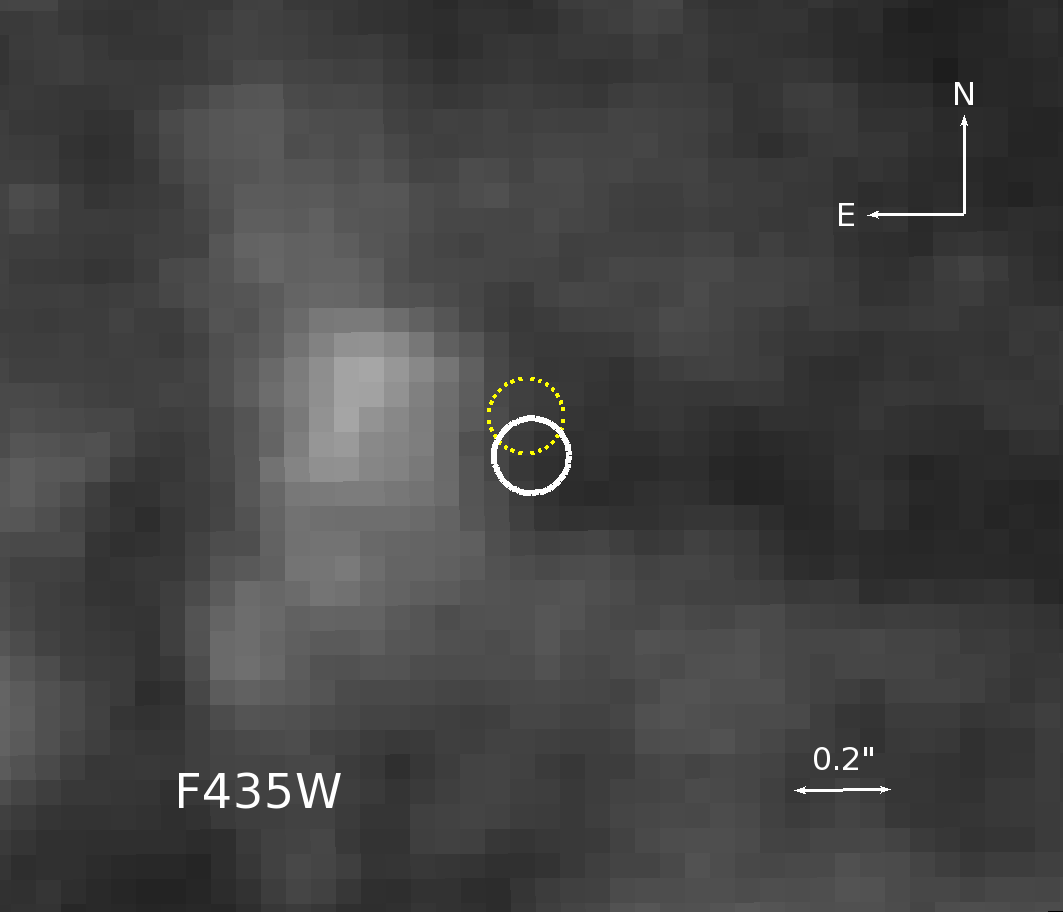}}
\subfigure{\includegraphics[angle=0,width=3.25in]{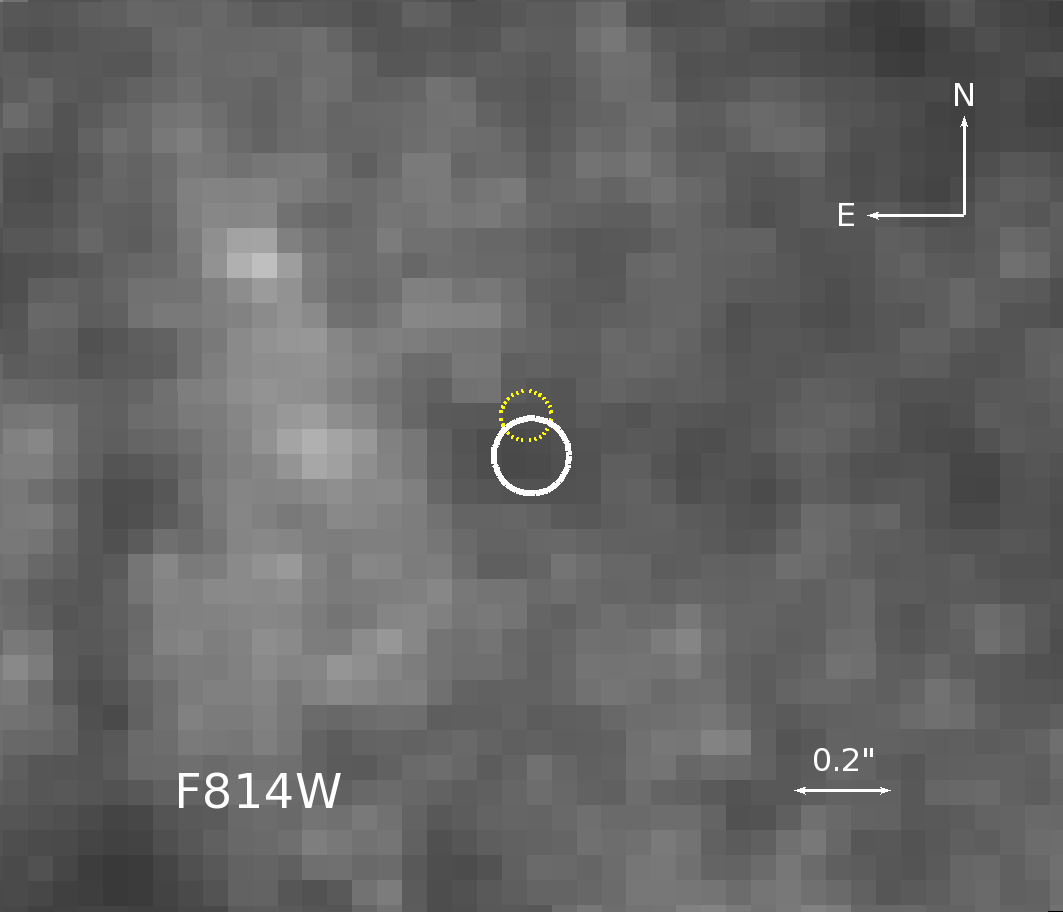}}
\subfigure{\includegraphics[angle=0,width=3.25in]{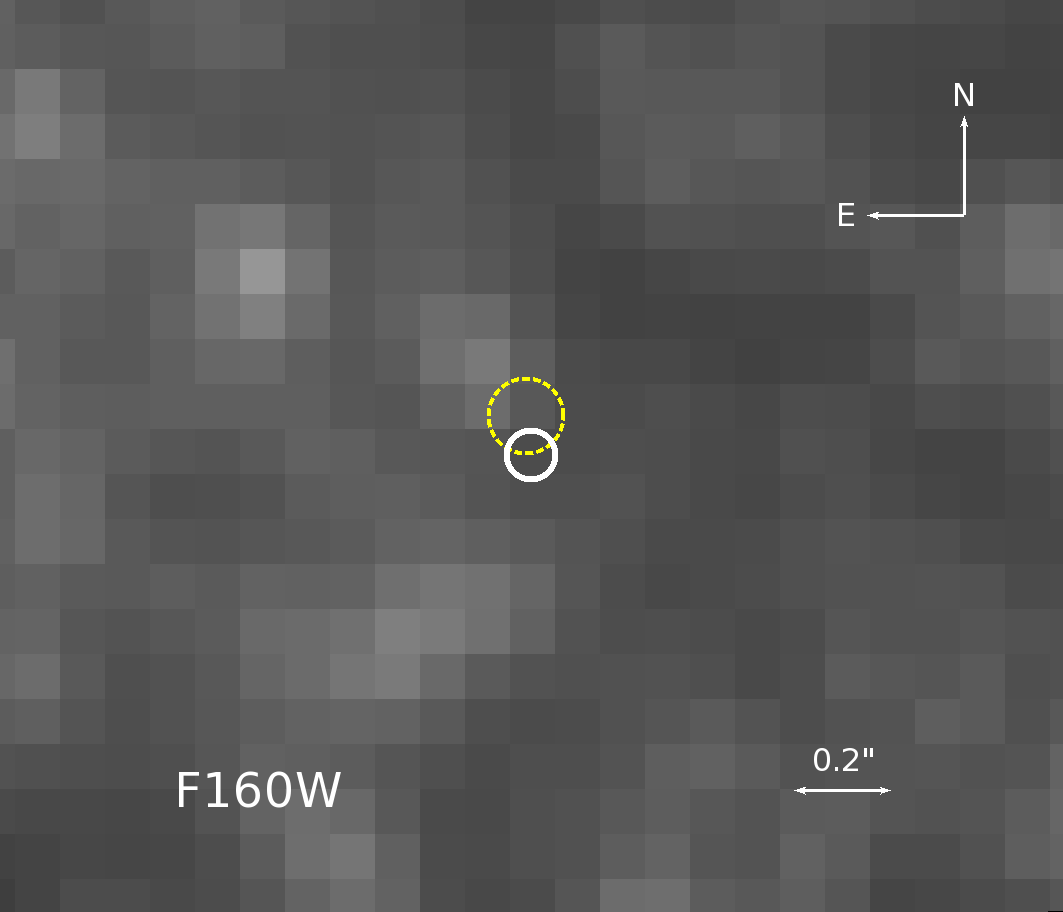}}
\caption{AO position of SN~2014J in a coadded image of all pre-explosion {\it HST} exposures, as well as in coadded F435W, F814W, and F160W {\it HST} images. 
The center of the solid white circle shows the position that we measure from our AO data, 
while the center of the dashed yellow circle corresponds to the position
published by Tendulkar et al. (\citeyear{tendulkarliu14}; also \citealt{goobarjohansson14}) from separate AO observations. 
The root-mean square (RMS) scatter of the astrometric fit between our NIRC2 AO image and the HLA F160W image
is \keckrarms~in RA and \keckdecrms~in Dec, and that reported by \citet{tendulkarliu14} relative to the HLA F814W image is \eightfourteenonesixtyrarms~in RA and \eightfourteenonesixtydecrms~in Dec.
The radii of the circles shown in each image correspond to the positional uncertainty of the SN in either the F160W or F814W image, respectively, convolved when appropriate with the RMS astrometric scatter between images in different bands (e.g., F160W and F435W). 
Our position is farther from a source considered as a possible progenitor candidate by \citet{goobarjohansson14} and is coincident with a region having strong extinction from dust. }
\label{fig:hstimages}
\end{figure*} 

\section{Discovery and Early Analysis of SN~2014J}
\label{sec:discovan}
SN~2014J was discovered by \citet{fosseycooke14} on \fosseyutdate~(UT dates are used throughout this paper) during a University College London class observing session.
\citet{zhengshivvers14} found that the light curve favors a time of first light of 
\zhengdiscov, and showed that the early evolution of the light curve 
can be best described by a varying power law that shows a more rapid initial rise in luminosity than a 
$t^2$ dependence. 

Tendulkar et al. (\citeyear{tendulkarliu14}; also \citealt{goobarjohansson14}) acquired {\it K}-band images with the Near Infrared Camera 2 (NIRC2) in conjunction with the adaptive optics (AO) system \citep{wizinowichlemignant06} on the Keck-II 10-m telescope. 
The corrected point-spread function (PSF) of their coadded image had a full width at half-maximum intensity (FWHM) of 0.36$''$ (measured at a 10$''$ offset from SN~2014J). 
\citet{goobarjohansson14} consider a source 0.2$''$ (5$\sigma$) from the measured SN position as a potential companion,
although they suggest that a radio upper limit \citep{chandlermarvil14}   
disfavors this possibility by setting a prohibitively low constraint on a donor star's mass-loss rate. 

After removing \MWebv~Galactic reddening determined by \citet{schlaflyfinkbeinerSFD11}, \citet{goobarjohansson14} find that the shape of the optical SN~2014J spectrum may be reproduced
by applying a low $R_V=A_V/E(B-V)$ extinction curve to a spectrum
of SN~2011fe, a spectroscopically normal and unreddened SN~Ia; 
a \citet{cardelli89} reddening law with \goobarebv~and \goobarRv~provides the best 
match to the observed SN~2014J spectrum.
The favored value of $A_V$ is consistent with the measured equivalent width (EW) of the diffuse interstellar band (DIB) at 5780~\AA.

\citet{nielsengilfanov14} analyze archival observations taken with the {\it Chandra} X-ray telescope of the
position of SN~2014J. They are not able to exclude a low-temperature ($kT_{\rm eff} \lesssim 80$~eV) 
supersoft X-ray source coincident with the
explosion. Additionally, they find that the explosion coordinates lie near the center of a 
$\sim200$~pc structure of diffuse X-ray emission with inferred mass $\sim3 \times 10^4$~M$_{\odot}$,
which they suggest may be a bubble inflated by one or more previous SN. \citet{nielsengilfanov14} 
consider the possibility that SN~2014J may be associated with a prompt channel linked to nearby recent star formation.

\section{Data}
\label{sec:data}

\subsection{Keck-II NIRC2 Adaptive Optics Imaging}

To locate the SN site in pre-explosion {\it HST} exposures with high precision, we acquired wide-field NIRC2 {\it K}-band AO images of the site of SN~2014J on \aoutdate. The SN was used as the guide star to measure tip-tilt corrections. The FWHM natural seeing produced a PSF with 0.65$''$ FWHM, while the AO-corrected images have a PSF FWHM of 0.1$''$ near the SN position.

We first subtracted the median of a stack of bias exposures and then applied flat-field and distortion\footnote{https://www2.keck.hawaii.edu/inst/nirc2/dewarp.html} corrections to the NIRC2 images. Although exposures of SN~2014J were acquired for the minimum possible integration time, 
the number of counts registered by the several pixels closest to the peak of the PSF  
exceeded the range within which the detector has a linear response.
Both to register the \aoexp~AO exposures and then to measure the 
coordinates of the PSF of SN~2014J, we needed to determine the center of the PSF of the SN. 
We therefore fit a two-dimensional Gaussian model to
the SN pixel intensity distribution in each of the \aoexp~exposures and allowed the PSF center coordinates, FWHM, and normalization to vary. 

Saturated pixels in the NIRC2 detector assume low data number (DN) values during readout. For each exposure, we identified (through visual inspection) and masked the one or two pixels whose low DN values were consistent with a saturated pixel. Our $\chi^2$ goodness-of-fit statistic excludes pixels where the model value was in excess of the nonlinearity threshold.
Centers of the fitted Gaussian models coincided with the positions of the most saturated pixel. 
The NIRC2 exposures show no evidence for bleeding from saturated pixels. 

We then used the best-fitting SN coordinates to align the reduced images.
The positions of the intensity peaks of many sources in M82 show variation with wavelength; thus, to mitigate the contribution of this astrometric noise and to maximize the number of common sources,
we registered the coadded {\it K}-band image against archival {\it HST} F160W ({\it H}) images. 
To minimize the effect of any remaining astrometric distortion in the NIRC2 coadded image, we 
used sources only within the central \aodim~section of the NIRC2 coadded image when cross-registering with the {\it HST} F160W image. 
After extracting sources with SExtractor \citep{bert96} and excluding extended objects with $r_{50} > 3$ pixels,
we used the {\tt tweakreg} routine in the {\tt astrodrizzle} package\footnote{http://drizzlepac.stsci.edu/} to fit for an astrometric solution from \keckstarmatches~common objects.

Figure~\ref{fig:AO} shows the astrometrically matched, coadded {\it K}-band exposure obtained with the NIRC2 AO system, along with
the coadded {\it HST} F160W image. 
The position of the SN is \sncoordsme~in the World Coordinate System (WCS) of the drizzled F160W image\footnote{hst\_11360\_r9\_wfc3\_ir\_f160w\_drz.fits} available from the Hubble Legacy Archive (HLA)\footnote{http://hla.stsci.edu/}.
The RMS of the offsets between pairs of matched objects after applying the astrometric solution is \keckrarms~in RA and \keckdecrms~in Dec. 

\subsection{Pre-Explosion HST Imaging}
Table~\ref{tab:datasets} provides a list of the {\it HST} imaging datasets of the explosion site that we use to 
place limits on the flux of the progenitor. 
Except for the Advanced Camera for Surveys (ACS) exposures, the images within each dataset were processed, drizzled,
and coadded to a common pixel grid by the HLA team. 
The ACS coadded mosaics of drizzled images that we analyze are High Level Science Products (HLSP) made available by 
the observers who acquired the data (Proposal 10776; PI: M.~Mountain).

\begin{figure*}[t]
\centering
\includegraphics[angle=0,width=7.5in]{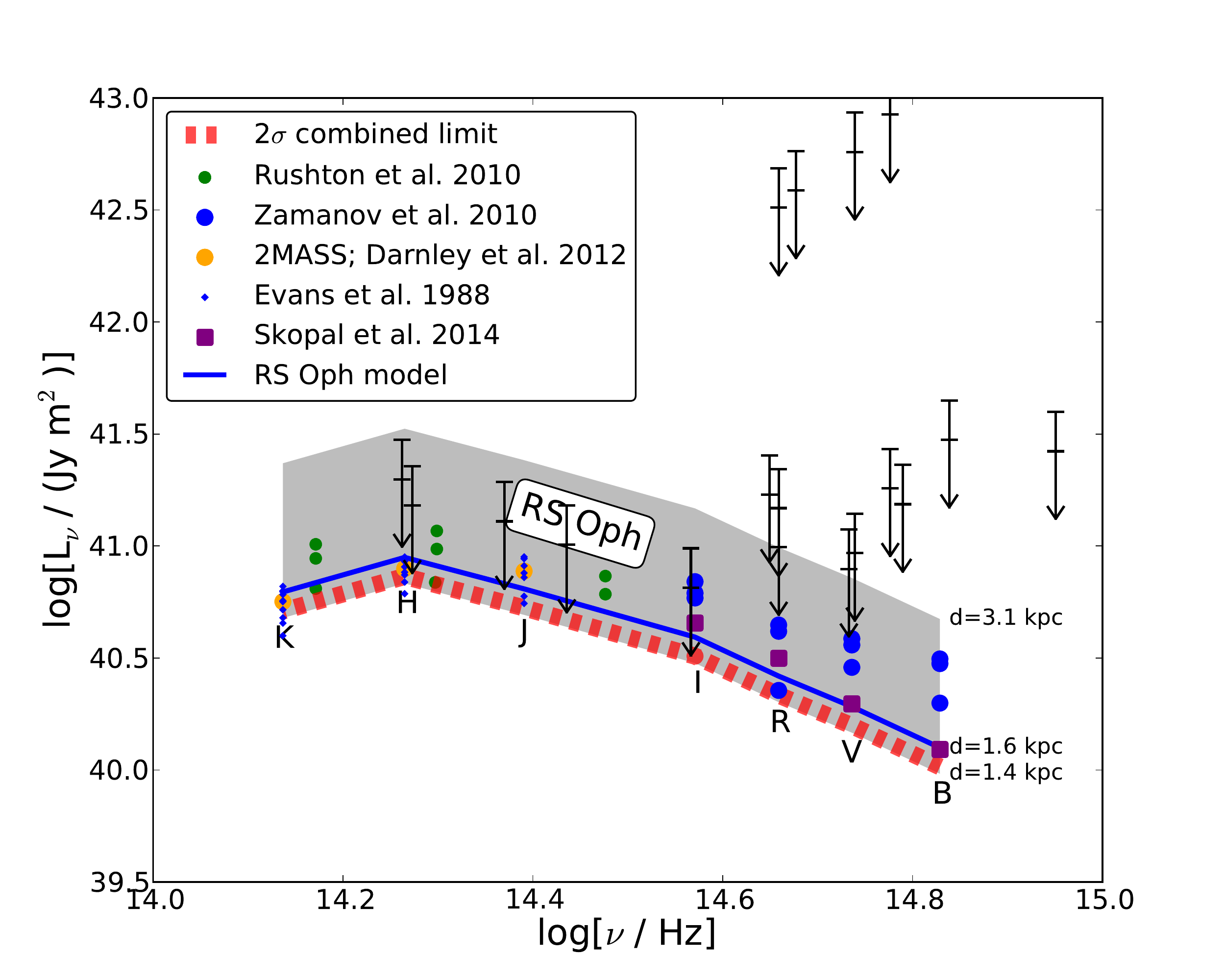}
\caption{Constraints on the luminosity $L_{\nu}$ of the progenitor system against the central frequency $\nu$ of the bandpass filter. Values for the Galactic symbiotic nova RS~Oph, a candidate single-degenerate progenitor system, in the quiescent phase are plotted for comparison. For each {\it HST} imaging dataset, the arrow tip marks the estimated 1$\sigma$ upper luminosity limit, while the horizontal hatch marks along the arrow show the 2$\sigma$ and 3$\sigma$ upper limits.  
The thick blue line shows the conservative values we adopt as a model for the luminosity of RS~Oph in the quiescent phase (with $d=1.6$~kpc), while the gray region shows the range in $L_{\nu}$ for the range of possible distances to RS~Oph ($1.4\leq d \leq3.4$~kpc).
The red dashed line represents the 2$\sigma$ limits inferred from {\it HST} observations in all bandpasses using 
the model SED for RS~Oph.
The progenitor luminosity limits shown are computed using the extinction and reddening \goobarext~values along the line of sight to SN~2014J estimated by \citet{goobarjohansson14}. 
We plot the central frequencies of the Johnson-Cousins {\it BVRI} and Bessell {\it JHK} filters for reference. 
Measurements of RS~Oph from \citet{evanscallus88}, \citet{rushtonkaminsky10}, \citet{zamanovboeva10}, \citet{darnleyribeiro12}, and \citet{skopal14} are converted to $L_{\nu}$ for $d=1.6$~kpc.
}
\label{fig:rsoph}
\end{figure*} 

\begin{figure*}[t]
\centering
\includegraphics[angle=0,width=7.5in]{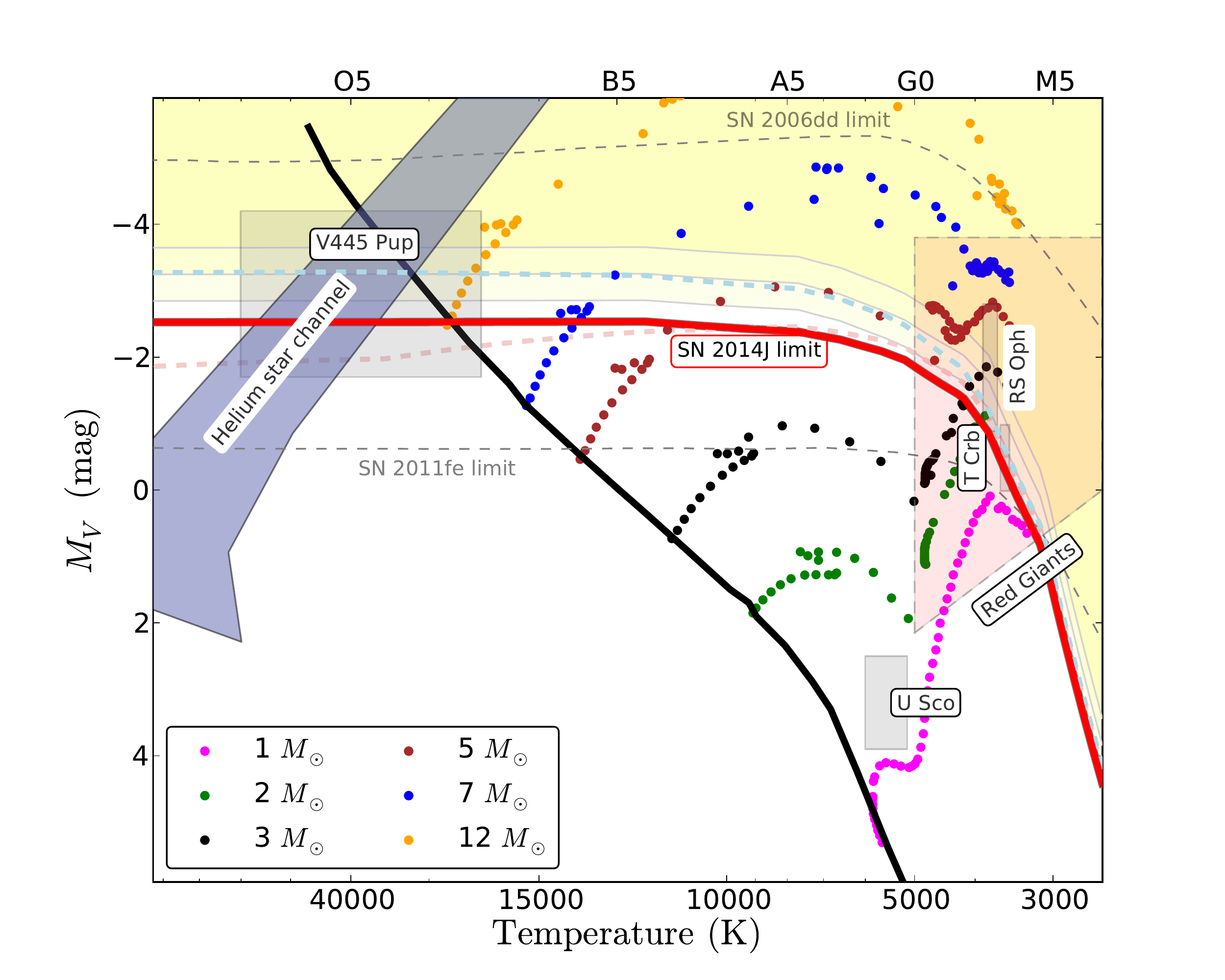}
\caption{Constraints on the position of the SN~2014J progenitor system in the Hertzsprung-Russell (H-R) diagram.  
The thick bright red line corresponds to the 2$\sigma$ $M_V$ limits \goobarext~that we place on the progenitor system as a function of temperature and stellar type from the combination of all limits in all {\it HST} bands. 
The middle solid gray line shows the 2$\sigma$ limits on the progenitor absolute magnitude obtained using the most constraining single observation, while the upper and lower adjacent gray lines provide an estimate of the uncertainty on this limit.
The region brighter than the most constraining single observation limit is shaded in yellow. 
Starburst regions of M82 have approximately solar metallicity \citep{forsterschriebergenzel01}, 
and we plot stellar evolutionary tracks off of the main sequence (solid black line) calculated by \citet{lejeuneschaerer01} with appropriate abundance ($Z = 0.02$).
Our upper limits likely exclude recurrent novae with luminosities comparable to that of RS~Oph \citep{hachisukato01}.
Candidate systems U~Sco \citep{thoroughgooddhillon01,hachisukato99} cannot be excluded, while T~CrB \citep{hachisukato01} and V445~Pup \citep{woudtsteeghs09} span the upper limit.
The light-red shaded region corresponds approximately to stars considered 
red giants and red bright giants.  
Dashed gray lines show the limits measured for SN~2011fe in M101 \citep{libloom11} as well as SN 2006dd in NGC 1316 \citep{maozmannucci08}
(representative of constraints for other nearby SN~Ia).
The dashed pale red line shows the  2$\sigma$ $M_V$ limits for \pBRv~and~\pBAv~mag, while the dashed pale blue line shows those for \pCRv~and~\pCAv~mag. 
Absolute magnitudes are computed using a distance modulus of \distmod.
}
\label{fig:Vband}
\end{figure*}

\section{Methods}
\label{sec:methods}
We find an astrometric solution that aligns each coadded, drizzled {\it HST} image (see Table~\ref{tab:datasets})
with the F160W reference coadded image through a sequence of two steps. 
The objective is to achieve an accurate astrometric solution near the coordinates of SN~2014J, so we 
trim each {\it HST} image to a \mbox{$30''\times30''$} subsection 
centered on the SN position.
In the first step, we use the IRAF\footnote{http://iraf.noao.edu/} {\tt ccmap} routine to compute an approximate astrometric alignment 
using $\sim5$--10 sources in common between the images. 
Next we use Source Extractor (SExtractor; \citealt{bert96}) to extract a 
catalog of objects in each image, and we derive a final astrometric solution from $\sim70$--500 sources
using the {\tt tweakreg} routine in the {\tt astrodrizzle} package. 

In Figure~\ref{fig:hstimages}, we show our measured SN coordinates, as well as those obtained by \citet{tendulkarliu14},
in a coadded image of all pre-explosion {\it HST} exposures, and the F435W (Johnson {\it B}), F814W (Wide {\it I}), and F160W ({\it H}) {\it HST} coadded images. 
The \citet{tendulkarliu14} position was reported relative to the WCS of the HLA F814W image, and we use our astrometric registration of the images to determine the location of the \citet{tendulkarliu14} position in our reference F160W image. 
The SN~2014J position that we measure is offset by~\offsetusatel~from the coordinates we calculate for the \citet{tendulkarliu14} F814W position in the F160W image.

The angular distance between the position we estimate and the preliminary coordinates reported by \citet{tendulkarliu14} may arise
from several differences between our AO coadded images and astrometric fitting. 
These include the substantially improved resolution of our NIRC2 AO exposures (0.1$''$) compared to those analyzed by Tendulkar et al. (\citeyear{tendulkarliu14}; 0.36$''$), our restriction of cross-matched sources to those inside of the central 16$''$~$\times$~16$''$ region of the 40$''$~$\times$~40$''$ 
wide-field NIRC2 camera to minimize the effects of residual distortion, the numbers of matched sources (\keckstarmatches~and 8, respectively) incorporated into the astrometric fit by the two analyses, and our matching of sources in the {\it K}-band NIRC2 image against the near-IR {\it HST} F160W image as opposed to the {\it I}-band F814W image to be able to minimize source confusion and the effects of differential reddening.

\subsection{Upper Flux Limits}
As may be seen in the representative images in Figure~\ref{fig:hstimages}, the local environment of SN~2014J exhibits strong surface brightness variations
from both resolved and unresolved sources, as well as strong and varying extinction by dust. 
As in \citet{maozmannucci08}, our approach for estimating nondetection limits is to add artificial point sources.
We create a sequence of images, each with a superimposed artificial point source having FWHM appropriate to the {\it HST} instrument and bandpass filter.
We create a pixelated Gaussian PSF and use the PyFITS package\footnote{http://www.stsci.edu/institute/software\_hardware/pyfits} to add the artificial source to the {\it HST} image.
We place the PSF $\sim1$--2 pixels in a randomly determined direction
away from the SN explosion site to avoid superimposing the PSF on any 
possible underlying counts from a progenitor, which would yield a biased upper limit.
These offset positions are sufficiently modest that the local
background is comparable to that at the SN site.  

The artificial point source in each successive image contains 0.1~mag additional flux than that in the previous image. 
We visually determine the limiting magnitude by identifying the first image for which the brightest pixel of the injected source 
has more counts than the neighboring local maxima produced by photon shot noise and background variation, 
and where pixels adjacent to the peak of the injected source have elevated counts consistent with a PSF.

To estimate the statistical significance of these visual detections for use in likelihood computations, 
we compute detection limits using a complementary technique.
We calculate the RMS of the background flux in a region without detected sources or a strong intensity gradient. 
We consider a 3$\sigma$ detection to be one for which the source flux exceeds the background RMS inside a 6-pixel aperture by a factor of 3. 

A circular aperture enclosing 80\% of the flux of a point source is the default 
choice for most {\it HST} instruments when computing the signal-to-noise ratio (S/N) of a flux measurement using 
the Exposure Time Calculator (ETC)\footnote{http://etc.stsci.edu/etc/}. 
For the ACS Wide Field Camera (WFC), this corresponds to an area that includes $\sim44$ pixels (and corresponding background noise), and yields a limiting magnitude that is $\sim0.7$ mag brighter and less sensitive than the value we compute 
for a 6-pixel background aperture. 
From inspection of the artificial point sources, we find that visual detections, however, depend almost entirely on only the several pixels closest to the coordinates of the source.
We note that the limiting magnitudes of {\it HST} images calculated by \citet{libloom11} in their analysis of the explosion site of SN~2011fe are consistent with estimates instead made using a circular aperture enclosing 80\% of the flux.

As shown in Table~\ref{tab:datasets}, the visual magnitude limits are 
comparable to the 3$\sigma$ limits estimated using the background statistics. 
We assign accordingly a 3$\sigma$ significance to the visual limits for the purpose of computing likelihood functions.

\subsection{Constraints on Progenitor Systems}

To convert these limits on the apparent magnitudes of a progenitor system to constraints
on a stellar source, we use the \citet{pickles98} library to model the spectral energy distributions (SEDs) of 
candidate progenitor systems of the SN, as well as (following \citealt{libloom11}) blackbody spectra with 
temperatures of 35,000, 65,000, and 95,000~K. 
For each combination of {\it HST} filter and instrument in Table~\ref{tab:datasets} and all spectroscopic templates,
we compute synthetic magnitudes with and without extinction from dust. 
After removing \MWebv~of foreground Milky Way ($R_V=3.1$) extinction, 
\citet{goobarjohansson14} estimated that additional extinction of \goobarebv~with 
\goobarRv~can best reproduce the shape of a premaximum optical spectrum of SN~2014J. 
While we adopt the \citet{goobarjohansson14}~extinction curve for our progenitor constraints, 
we additionally compute synthetic magnitudes instead with an $R_V=3.1$ curve for M82, using the total $A_V$ favored by 
\citet{goobarjohansson14}.
As we demonstrate in \S \ref{sec:results}, we obtain similar progenitor constraints for these differing
values of $R_V$.

Following \citet{libloom11}, we first translate the upper limit on the flux for each filtered observation to 
a 2$\sigma$ upper limit on the absolute magnitude $M_V$, using a distance modulus to M82 of \distmod.
After repeating this for all observations, we 
identify the most constraining upper limit on $M_V$, and refer to this as the 2$\sigma$ ``1-frame'' limit for each spectrum template. A combined constraint is estimated next by constructing a probability function that incorporates the upper magnitude limits in all filters.
The total probability (as a function of $M_V$ and the template spectrum) is the product of the probabilities of each filtered observation computed from the predicted model magnitude and the measured upper magnitude limit. 

To calculate the combined progenitor system limit, we incrementally increase the absolute brightness $M_V$ until a 2$\sigma$ probability ($p=0.954$) is reached.
As in \citet{libloom11}, the combined constraints are $\sim0.2$--0.8 mag deeper than ``1-frame'' constraints which rely on observations in a single filter.  We would expect to detect, with 95\% probability, such a 2$\sigma$ source in a least one {\it HST} bandpass image, and in the coadded image in Figure~\ref{fig:hstimages}.

Table~\ref{tab:datasets} shows the detection upper limits for the coadded image of each dataset, 
and Table~\ref{tab:hrlimits} lists the corresponding constraint on $M_J$, in addition to $M_V$, 
computed for each \citet{pickles98} and blackbody spectrum.  
For each spectrum, the photometric band providing the faintest 
individual absolute magnitude limit is also identified in Table~\ref{tab:hrlimits}.

In Figure~\ref{fig:rsoph}, we plot the limits on the progenitor flux we estimate using the
{\it HST} imaging. We also show measurements of the luminosity $L_{\nu}$ of RS~Oph 
during its quiescent phase as a function of frequency $\nu$,
and the uncertainty arising from current constraints on its distance.

\subsection{RS~Oph SED Model}
\label{sec:rsoph}

The luminosity of RS~Oph is comparable to our upper detection limit \goobarext~mag, and the bright white dwarf primary in RS~Oph contributes significantly in bluer
optical bandpasses, unlike the primary of T~CrB. Therefore, we create a near-UV to near-IR model of the spectrum of RS~Oph to account for the contribution of the white dwarf, and to show clearly the measurements that contribute to our constraints.

\citet{anupamamikolajewska99} performed a decomposition of an optical spectrum of RS~Oph in its quiescent phase
and find an A2 to A4 stellar spectrum for the hot component, and an M0 to M2 III spectrum for
the giant companion. We fit for the linear combination of the corresponding \citet{pickles98} stellar spectra that best matches the
broadband magnitudes of RS~Oph during quiescence.   \citet{libloom11} use an effective temperature of 3750 to 3650 K for the giant companion typical of  M0 to M2 giant stars \citep{straizyskuriliene81} in their SN~2011fe progenitor analysis.

\citet{evanscallus88} compiled {\it J}-, {\it H}-, and {\it K}-band magnitudes of RS~Oph obtained during quiescence from 1971 through 1982 by \citet{swingsallen72}, \citet{feastglass74}, \citet{szkody77}, \citet{sherringtonjameson83}, and \citet{kenyongallagher83}. These measurements show modest variation of $\pm$0.2 mag, and 
we use the median values ($J\approx7.67$, $H\approx6.92$, $K\approx6.62$ mag) as representative values
during the quiescent phase. Magnitudes of RS~Oph measured by 2MASS \citep{skru06} in 1999 ($J=7.63\pm0.02$, $H=6.85\pm0.04$, $K=6.50\pm0.01$ mag; see also \citealt{darnleyribeiro12}), as well as values synthesized from a near-IR spectrum \citep{rushtonkaminsky10,skopal14}, are consistent with the magnitudes in quiescence assembled by \citet{evanscallus88}.
We use the optical {\it BVRI} photometry of RS~Oph during quiescence measured by \citet{skopal14} to construct our 
model. 
The \citet{skopal14} optical measurements are, on average, fainter than those presented by \citet{zamanovboeva10} for several
epochs. 
While \citet{schaefer10} combine optical and IR colors of RS~Oph taken on two dates to infer a possibly fainter flux in the IR, the combined values they present do not appear to be representative of the system in quiescence.

To estimate the luminosity of RS~Oph, we use \ebvrsoph~mag \citep{snijders87}, which is consistent with the value derived by \citet{anupamamikolajewska99}, for an $R_V=3.1$ \citet{cardelli89} Milky Way extinction curve.
Estimates for the distance to RS~Oph include $\sim1.4$~kpc from modeling of the 2006 outburst (e.g., \citealt{rupenmioduszewski08}), $\sim1.6$~kpc from radio observations of expanding material in the 1985 \citep{hjellmingvangorkum86}  and 2006 \citep{sokoloskiluna06} outbursts, and $\sim3.1$~kpc based on the assumption that the secondary fills the entire Roche lobe \citep{liviotruran86,barrymukai08}.

With a {\it V}-band total magnitude of 11.5 in quiescence, we estimate a range in absolute magnitude of $-1.49 \leq M_V \leq -3.22$
for distances of 1.4~kpc to 3.1~kpc. 
Using our decomposition of the spectrum of RS~Oph in quiescence, we estimate that the giant star should be $\sim0.65$ mag fainter in $V$ than the total magnitude of the binary system.

\section{Results}
\label{sec:results}
\subsection{Progenitor Surroundings}
As can be seen in Figure~\ref{fig:hstimages}, the SN~2014J coordinates that the AO analysis
favors are at a greater angular distance from the nearest resolved point source than 
the \citet{tendulkarliu14} position. This suggests that the star at a smallest angular offset is not a mass donor to the 
white dwarf progenitor of SN~2014J.  Our coordinates also position SN~2014J closer to the 
center of an apparent dust cloud whose silhouette is visible in optical {\it HST} images. 

\subsection{Progenitor Models}
Figure~\ref{fig:Vband} shows the derived constraints on the position of the progenitor system of 
SN~2014J in the Hertzsprung-Russell (H-R) diagram. 
The upper luminosity limit excludes the bright extent of 
the region, marked in pale red, occupied by red giant stars.
The faint boundary of the red giant region corresponds approximately to the least luminous 
red giants in the Hipparcos catalog \citep{perrymanlindegren97}, and
we truncate the red giant region at an upper effective temperature of 5000~K.
Stars more luminous than the plotted $M_V \approx 3.8$~mag bright boundary may instead be classified as 
red supergiant stars. 
 
With the exception of values for RS~Oph (see \S \ref{sec:rsoph}), we plot the ranges of effective temperature and $M_V$ values collected by \citet{libloom11} for Galactic candidate systems. 
Our luminosity limits are comparable to our faintest estimates 
for the luminosity of RS~Oph in quiescence, including the uncertainty in its distance in the Galaxy. 
The upper $M_V$ line intersects the $M_V$--$T$ area corresponding to the less luminous Galactic symbiotic system T~CrB \citep{hachisukato01}.

U~Sco is a recurrent nova and supersoft X-ray source with a large white
dwarf mass of $1.55 \pm 0.24$~M$_{\odot}$ \citep{thoroughgooddhillon01} and a
subgiant companion. This candidate progenitor system is substantially too faint for
detection in our archival images, as was the case for SN~2011fe. 

A candidate He-rich single-degenerate progenitor system is the He 
nova V445 Puppis (V445~Pup). We use the \citet{woudtsteeghs09} estimate for $M_V$ based on a parallax distance 
measurement of $8.2 \pm 0.5$~kpc from the expansion of a bipolar shell 
after an eruption in year 2000. The upper luminosity limits for SN~2014J extend across the region of the H-R diagram 
corresponding to V445~Pup. 
Figure~\ref{fig:Vband} shows the theoretical limits on the He channel 
computed by \citet{libloom11} using \citet{liuchen10} models and bolometric corrections 
from \citet{torres10}. 
Our measured limits are fainter than the predicted luminosities of He-star-channel 
progenitors with comparatively cool temperatures ($T \lesssim$ 35,000~K).

For comparison, we plot the luminosity limits for the progenitors of SN~2011fe \citep{libloom11} and 
SN~2006dd \citep{maozmannucci08}; the latter provides a constraint typical of SN~Ia other than SN~2011fe. 
For companions with very low effective temperature ($T < 3000$~K), 
limits on the progenitor SN~2014J are significantly fainter than those for SN~2011fe,
because the SN~2014J archival images extend to the near-IR.

Now referring to the stellar evolutionary tracks off the main sequence computed 
by \citet{lejeuneschaerer01}, we find that a mass donor with effective temperature
cooler than 4000~K would need to have a mass less than 2~M$_{\odot}$. 
The plotted evolutionary tracks are models with abundance ($Z = 0.02$)
consistent with the metallicity of M82 starburst regions \citep{forsterschriebergenzel01}.
The derived luminosity limits are significantly brighter than a hypothetical binary system of two white dwarfs that does not experience a long-lived merger phase \citep{shenbildsten12}, or a white-dwarf/main-sequence binary (e.g., \citealt{wheeler12}). 

Comparison among the bright red (\pARv; \pAAv~mag; \citealt{goobarjohansson14}), the dashed pale red (\pBRv; \pBAv~mag),
and the dashed pale blue (\pCRv; \pCAv~mag) lines shows that the progenitor constraints exhibit only modest change within 
the range of dust parameters reported from analyses of the SN~2014J light curve and spectra (e.g., \citealt{goobarjohansson14}; \citealt{patattaubenberger14}). 

While the total integration times of pre-explosion {\it HST} broadband imaging of the site of SN~2014J are almost all greater than that of the site of SN~2011fe, and M82 ($d \approx 3.5$;~\citealt{jacobsrizzi09}) is at a smaller distance than M101 ($d \approx 6.4$~Mpc; \citealt{shappeestanek11}), 
the upper flux limits at optical wavelengths that we estimate for the progenitor of SN~2014J are less constraining than those for the progenitor of SN~2011fe \citep{libloom11}. The brighter optical constraints on the progenitor of SN~2014J arise from the high extinction inferred along the line of sight to SN~2014J, and from the bright and varying local background in M82; the spectra and colors of SN~2011fe were consistent with a lack of extinction and it occurred at a location with low surface brightness. In the ACS WFC F814W FLT images of the explosion sites, we measure an average of $\sim$0.13~counts~s$^{-1}$~pix$^{-1}$ close
to the explosion site of SN~2011fe in M101, and $\sim$5.3~counts~s$^{-1}$~pix$^{-1}$ close to SN~2014J in M82.

\section{Conclusions}
\label{sec:conclusions}
We have used archival, pre-explosion {\it HST} images of M82 in the near-UV through 
near-IR to place constraints on the progenitor system of the Type~Ia SN~2014J. 
Assuming that the extinction and selective extinction along the line of sight to the SN 
estimated from the SN light curve and optical spectra are approximately correct (e.g., \citealt{goobarjohansson14}; \citealt{patattaubenberger14}),
we can exclude a progenitor system with a bright red giant mass-donor companion, including recurrent novae with luminosities comparable to the Galactic prototype symbiotic system RS~Oph.
Our limits are fainter than the predicted luminosity of He-star-channel progenitors with comparatively low effective temperature.
The available near-IR M82 data provide a fainter limit for mass donors with very low 
effective temperatures ($T < 3000$~K) than was possible at the 
explosion site of SN~2011fe in M101. 
A hypothetical progenitor system consisting of two white dwarf stars that does not experience a long-lived merger phase \citep{shenbildsten12} 
would have a luminosity significantly fainter than the upper limits we estimate. 

\begin{deluxetable*}{cccccccc}
\tablecaption{{\it HST} Datasets and Upper Absolute Magnitude Limits on Point-Source Flux at Explosion Site}
\tablecolumns{8}
\tablehead{ \colhead{Instrument}&\colhead{Aperture}& \colhead{Filter}&\colhead{UT Date Obs.}&\colhead{Exp. Time (s)}&\colhead{Prop. No.}& \colhead{Visual Limit}& \colhead{3$\sigma$ Background Limit} }
\startdata
WFC3&UVIS&F225W&2010-01-01&1665.0&11360&26.50&26.80\\
WFC3&UVIS&F336W&2010-01-01&1620.0&11360&26.71&27.23\\
ACS&WFC&F435W&2006-09-29&1800.0&10766&26.30&27.05\\
WFC3&UVIS&F487N&2009-11-17&2455.0&11360&26.01&25.94\\
WFC3&UVIS&F502N&2009-11-17&2465.0&11360&25.93&26.28\\
WFPC2&WF&F502N&1998-08-28&3600.0&6826&21.76&22.70\\
WFC3&UVIS&F547M&2010-01-01&1070.0&11360&26.14&25.94\\
WFPC2&WF&F547M&1998-08-28&100.0&6826&21.63&22.12\\
ACS&WFC&F555W&2006-03-29&1360.0&10766&26.42&26.52\\
WFPC2&WF&F631N&1998-08-28&1200.0&6826&21.43&22.17\\
ACS&WFC&F658N&2004-02-09&700.0&9788&24.63&24.76\\
ACS&WFC&F658N&2006-03-29&4440.0&10766&25.06&25.17\\
WFPC2&WF&F658N&1997-03-16&1200.0&6826&21.31&21.86\\
WFC3&UVIS&F673N&2009-11-15&2760.0&11360&24.53&25.62\\
ACS&WFC&F814W&2006-03-29&700.0&10766&24.83&25.09\\
WFC3&IR&F110W&2010-01-01&1195.39&11360&23.54&23.51\\
WFC3&IR&F128N&2009-11-17&1197.69&11360&22.90&22.85\\
WFC3&IR&F160W&2010-01-01&2395.39&11360&22.43&22.48\\
WFC3&IR&F164N&2009-11-17&2397.7&11360&21.98&22.17
\enddata
\tablecomments{ Limiting magnitudes in the Vega system for point sources near the explosion coordinates in the {\it HST} images. Visual limiting magnitudes are estimated by injecting a point source of increasing brightness in close proximity to the AO explosion coordinates, and identifying when a source is clearly detected. The 3$\sigma$ background detections are computed using the RMS of the background measured in a region without point sources or pronounced background gradients. }
\label{tab:datasets}
\end{deluxetable*}
\begin{deluxetable*}{cccccc}
\tablecaption{Stellar and Blackbody Upper Magnitude Limits}
\tablecolumns{6}
\tablehead{                &\multicolumn{2}{c}{$M_V$ (2$\sigma$)}   &   \multicolumn{2}{c}{$M_J$ (2$\sigma$)}                & \colhead{Most Constraining} \\ 
\colhead{Star} & \colhead{``1-Frame''}& \colhead{Combined} & \colhead{`1'-Frame}& \colhead{Combined}&\colhead{`1'-Frame Bandpass}}
\startdata
O5 V&-3.25&-2.53&-2.51&-1.79&F555W\\
B0 V&-3.25&-2.55&-2.55&-1.85&F555W\\
A0 V&-3.26&-2.54&-3.26&-2.54&F555W\\
A5 V&-3.26&-2.48&-3.54&-2.76&F555W\\
F0 V&-3.08&-2.37&-3.61&-2.90&F814W\\
F5 V&-2.94&-2.26&-3.76&-3.08&F814W\\
G0 V&-2.76&-2.15&-3.77&-3.16&F814W\\
G5 V&-2.69&-2.08&-3.87&-3.26&F814W\\
K0 V&-2.55&-1.95&-3.93&-3.33&F814W\\
K5 V&-2.04&-1.40&-4.21&-3.57&F814W\\
M0 V&-1.63&-0.86&-4.49&-3.72&F110W\\
M4 V&-0.09&0.57&-4.52&-3.86&F110W\\
M5 V&0.71&1.43&-4.55&-3.83&F160W\\
B5 III&-3.25&-2.54&-2.92&-2.21&F555W\\
G0 III&-2.69&-2.03&-4.00&-3.34&F814W\\
G5 III&-2.55&-1.90&-4.09&-3.44&F814W\\
K0 III&-2.44&-1.79&-4.12&-3.47&F814W\\
K5 III&-1.81&-1.03&-4.48&-3.70&F110W\\
M0 III&-1.67&-0.86&-4.48&-3.67&F110W\\
M5 III&0.18&0.90&-4.48&-3.76&F160W\\
M10 III&3.84&4.42&-4.51&-3.93&F128N\\
B5 I&-3.26&-2.51&-3.09&-2.34&F555W\\
F0 I&-3.11&-2.38&-3.55&-2.82&F814W\\
F5 I&-3.04&-2.33&-3.70&-2.99&F814W\\
G0 I&-2.76&-2.13&-3.80&-3.17&F814W\\
G5 I&-2.60&-1.99&-3.91&-3.30&F814W\\
M2 I&-1.10&-0.42&-4.40&-3.72&F814W\\
BB1&-3.25&-2.53&-2.62&-1.90&F555W\\
BB2&-3.25&-2.53&-2.50&-1.78&F555W\\
BB3&-3.25&-2.53&-2.45&-1.73&F555W
\enddata
\tablecomments{Limiting magnitudes in $V$ and $J$ bands in the Vega system for a point source at the explosion site. The BB1, BB2, and BB3 blackbody spectra have 35,000, 65,000, and 95,000 K temperatures, respectively. Stellar classifications are those of the \citet{pickles98} spectra used as models of the potential companion. The bandpass in right column is the most constraining observation for the ``1-frame'' upper magnitude limits.  }
\label{tab:hrlimits}
\end{deluxetable*}

\acknowledgements
We greatly appreciate the critical contribution of observing time from Vithal Tilvi, Nicola Mehrtens,
Casey Papovich, and Mark Dickinson that enabled us to complete the Keck AO imaging, as well as their encouragement and helpful
suggestions on the manuscript. 
We thank Michael Liu, Shriharsh Tendulkar, Yi Cao, and Trent Dupuy for generously providing details 
of their NIRC2 AO observations to aid our team in deciding whether any further imaging could be helpful.
Correspondence with both Matthew Darnley and G. C. Anupama was also useful.
A.V.F.'s group at UC Berkeley has received generous financial assistance
from the Christopher R. Redlich Fund, the TABASGO Foundation, Gary and Cynthia Bengier, 
and NSF grant AST-1211916. This work was also supported by NASA grants AR-12623 and
AR-12850 from the 
Space Telescope Science Institute, which is operated by AURA,        
Inc., under NASA contract NAS 5-26555.
K.J.S. is supported by NASA through Einstein Postdoctoral Fellowship grant number PF1-120088 awarded by the Chandra X-ray Center, which is operated by the Smithsonian Astrophysical Observatory for NASA under contract NAS8-03060.

Some of the data presented herein were obtained at the W. M. Keck Observatory from telescope time allocated to NASA through the agency's scientific partnership with the California Institute of Technology and the University of California. The Observatory was made possible by the generous financial support of the W. M. Keck Foundation.  We recognize the Hawaiian community for the opportunity to conduct these observations from the summit of Mauna Kea.


\begin{thebibliography}{}
\expandafter\ifx\csname natexlab\endcsname\relax\def\natexlab#1{#1}\fi

\bibitem[{{Anupama} \& {Miko{\l}ajewska}(1999)}]{anupamamikolajewska99}
{Anupama}, G.~C., \& {Miko{\l}ajewska}, J. 1999, \aap, 344, 177

\bibitem[{{Barry} {et~al.}(2008){Barry}, {Mukai}, {Sokoloski}, {Danchi},
  {Hachisu}, {Evans}, {Gehrz}, \& {Mikolajewska}}]{barrymukai08}
{Barry}, R.~K., {Mukai}, K., {Sokoloski}, J.~L., {et~al.} 2008, in Astronomical
  Society of the Pacific Conference Series, Vol. 401, RS Ophiuchi (2006) and
  the Recurrent Nova Phenomenon, ed. A.~{Evans}, M.~F. {Bode}, T.~J. {O'Brien},
  \& M.~J. {Darnley}, 52

\bibitem[{{Bertin} \& {Arnouts}(1996)}]{bert96}
{Bertin}, E., \& {Arnouts}, S. 1996, AJ, 117, 393

\bibitem[{{Cardelli} {et~al.}(1989){Cardelli}, {Clayton}, \&
  {Mathis}}]{cardelli89}
{Cardelli}, J.~A., {Clayton}, G.~C., \& {Mathis}, J.~S. 1989, \apj, 345, 245

\bibitem[{{Chandler} \& {Marvil}(2014)}]{chandlermarvil14}
{Chandler}, C.~J., \& {Marvil}, J. 2014, The Astronomer's Telegram, 5812, 1

\bibitem[{{Childress} {et~al.}(2013){Childress}, {Aldering}, {Antilogus},
  {Aragon}, {Bailey}, {Baltay}, {Bongard}, {Buton}, {Canto}, {Cellier-Holzem},
  {Chotard}, {Copin}, {Fakhouri}, {Gangler}, {Guy}, {Hsiao}, {Kerschhaggl},
  {Kim}, {Kowalski}, {Loken}, {Nugent}, {Paech}, {Pain}, {Pecontal}, {Pereira},
  {Perlmutter}, {Rabinowitz}, {Rigault}, {Runge}, {Scalzo}, {Smadja}, {Tao},
  {Thomas}, {Weaver}, \& {Wu}}]{childressaldering13}
{Childress}, M., {Aldering}, G., {Antilogus}, P., {et~al.} 2013, \apj, 770, 108

\bibitem[{{Darnley} {et~al.}(2012){Darnley}, {Ribeiro}, {Bode}, {Hounsell}, \&
  {Williams}}]{darnleyribeiro12}
{Darnley}, M.~J., {Ribeiro}, V.~A.~R.~M., {Bode}, M.~F., {Hounsell}, R.~A., \&
  {Williams}, R.~P. 2012, \apj, 746, 61

\bibitem[{{Evans} {et~al.}(1988){Evans}, {Callus}, {Albinson}, {Whitelock},
  {Glass}, {Carter}, \& {Roberts}}]{evanscallus88}
{Evans}, A., {Callus}, C.~M., {Albinson}, J.~S., {et~al.} 1988, \mnras, 234,
  755

\bibitem[{{Feast} \& {Glass}(1974)}]{feastglass74}
{Feast}, M.~W., \& {Glass}, I.~S. 1974, \mnras, 167, 81

\bibitem[{{F{\"o}rster Schreiber} {et~al.}(2001){F{\"o}rster Schreiber},
  {Genzel}, {Lutz}, {Kunze}, \& {Sternberg}}]{forsterschriebergenzel01}
{F{\"o}rster Schreiber}, N.~M., {Genzel}, R., {Lutz}, D., {Kunze}, D., \&
  {Sternberg}, A. 2001, \apj, 552, 544

\bibitem[{{Fossey} {et~al.}(2014){Fossey}, {Cooke}, {Pollack}, {Wilde}, \&
  {Wright}}]{fosseycooke14}
{Fossey}, J., {Cooke}, B., {Pollack}, G., {Wilde}, M., \& {Wright}, T. 2014,
  Central Bureau Electronic Telegrams, 3792, 1

\bibitem[{{Geier} {et~al.}(2013){Geier}, {Marsh}, {Wang}, {Dunlap}, {Barlow},
  {Schaffenroth}, {Chen}, {Irrgang}, {Maxted}, {Ziegerer}, {Kupfer},
  {Miszalski}, {Heber}, {Han}, {Shporer}, {Telting}, {G{\"a}nsicke},
  {{\O}stensen}, {O'Toole}, \& {Napiwotzki}}]{geiermarsh13}
{Geier}, S., {Marsh}, T.~R., {Wang}, B., {et~al.} 2013, \aap, 554, A54

\bibitem[{{Goobar} {et~al.}(2014){Goobar}, {Johansson}, {Amanullah}, {Cao},
  {Perley}, {Kasliwal}, {Ferretti}, {Nugent}, {Harris}, {Gal-Yam}, {Ofek},
  {Tendulkar}, {Dennefeld}, {Valenti}, {Arcavi}, {Banerjee}, {Venkataraman},
  {Joshi}, {Ashok}, {Cenko}, {Diaz}, {Fremling}, {Horesh}, {Howell},
  {Kulkarni}, {Papadogiannakis}, {Petrushevska}, {Sand}, {Sollerman},
  {Stanishev}, {Bloom}, {Surace}, {Dupuy}, \& {Liu}}]{goobarjohansson14}
{Goobar}, A., {Johansson}, J., {Amanullah}, R., {et~al.} 2014, ArXiv e-prints,
  arXiv:1402.0849

\bibitem[{{Hachisu} \& {Kato}(2001)}]{hachisukato01}
{Hachisu}, I., \& {Kato}, M. 2001, \apj, 558, 323

\bibitem[{{Hachisu} {et~al.}(1999){Hachisu}, {Kato}, {Nomoto}, \&
  {Umeda}}]{hachisukato99}
{Hachisu}, I., {Kato}, M., {Nomoto}, K., \& {Umeda}, H. 1999, \apj, 519, 314

\bibitem[{{Han} \& {Podsiadlowski}(2004)}]{hanpodsiadlowski04}
{Han}, Z., \& {Podsiadlowski}, P. 2004, \mnras, 350, 1301

\bibitem[{{Hillebrandt} \& {Niemeyer}(2000)}]{hill00}
{Hillebrandt}, W., \& {Niemeyer}, J.~C. 2000, \araa, 38, 191

\bibitem[{{Hjellming} {et~al.}(1986){Hjellming}, {van Gorkom}, {Taylor},
  {Sequist}, {Padin}, {Davis}, \& {Bode}}]{hjellmingvangorkum86}
{Hjellming}, R.~M., {van Gorkom}, J.~H., {Taylor}, A.~R., {et~al.} 1986, \apjl,
  305, L71

\bibitem[{{Iben} \& {Tutukov}(1984)}]{ibe84}
{Iben}, Jr., I., \& {Tutukov}, A.~V. 1984, \apjs, 54, 335

\bibitem[{{Jacobs} {et~al.}(2009){Jacobs}, {Rizzi}, {Tully}, {Shaya},
  {Makarov}, \& {Makarova}}]{jacobsrizzi09}
{Jacobs}, B.~A., {Rizzi}, L., {Tully}, R.~B., {et~al.} 2009, \aj, 138, 332

\bibitem[{{Kasen} \& {Plewa}(2005)}]{kas05}
{Kasen}, D., \& {Plewa}, T. 2005, \apjl, 622, L41

\bibitem[{{Kasen} {et~al.}(2009){Kasen}, {R{\"o}pke}, \& {Woosley}}]{kas09}
{Kasen}, D., {R{\"o}pke}, F.~K., \& {Woosley}, S.~E. 2009, \nat, 460, 869

\bibitem[{{Kasen} \& {Woosley}(2007)}]{kas07}
{Kasen}, D., \& {Woosley}, S.~E. 2007, \apj, 656, 661

\bibitem[{{Kelly} {et~al.}(2010){Kelly}, {Hicken}, {Burke}, {Mandel}, \&
  {Kirshner}}]{kel10}
{Kelly}, P.~L., {Hicken}, M., {Burke}, D.~L., {Mandel}, K.~S., \& {Kirshner},
  R.~P. 2010, \apj, 715, 743

\bibitem[{{Kelly} {et~al.}(2008){Kelly}, {Kirshner}, \& {Pahre}}]{kel08}
{Kelly}, P.~L., {Kirshner}, R.~P., \& {Pahre}, M. 2008, \apj, 687, 1201

\bibitem[{{Kenyon} \& {Gallagher}(1983)}]{kenyongallagher83}
{Kenyon}, S.~J., \& {Gallagher}, J.~S. 1983, \aj, 88, 666

\bibitem[{{Lampeitl} {et~al.}(2010){Lampeitl}, {Smith}, {Nichol}, {Bassett},
  {Cinabro}, {Dilday}, {Foley}, {Frieman}, {Garnavich}, {Goobar}, {Im}, {Jha},
  {Marriner}, {Miquel}, {Nordin}, {{\"O}stman}, {Riess}, {Sako}, {Schneider},
  {Sollerman}, \& {Stritzinger}}]{lampeitl10}
{Lampeitl}, H., {Smith}, M., {Nichol}, R.~C., {et~al.} 2010, \apj, 722, 566

\bibitem[{{Lejeune} \& {Schaerer}(2001)}]{lejeuneschaerer01}
{Lejeune}, T., \& {Schaerer}, D. 2001, \aap, 366, 538

\bibitem[{{Li} {et~al.}(2011){Li}, {Bloom}, {Podsiadlowski}, {Miller}, {Cenko},
  {Jha}, {Sullivan}, {Howell}, {Nugent}, {Butler}, {Ofek}, {Kasliwal},
  {Richards}, {Stockton}, {Shih}, {Bildsten}, {Shara}, {Bibby}, {Filippenko},
  {Ganeshalingam}, {Silverman}, {Kulkarni}, {Law}, {Poznanski}, {Quimby},
  {McCully}, {Patel}, {Maguire}, \& {Shen}}]{libloom11}
{Li}, W., {Bloom}, J.~S., {Podsiadlowski}, P., {et~al.} 2011, \nat, 480, 348

\bibitem[{{Liu} {et~al.}(2010){Liu}, {Chen}, {Wang}, \& {Han}}]{liuchen10}
{Liu}, W.-M., {Chen}, W.-C., {Wang}, B., \& {Han}, Z.~W. 2010, \aap, 523, A3

\bibitem[{{Livio} {et~al.}(1986){Livio}, {Truran}, \&
  {Webbink}}]{liviotruran86}
{Livio}, M., {Truran}, J.~W., \& {Webbink}, R.~F. 1986, \apj, 308, 736

\bibitem[{{Maoz} \& {Mannucci}(2008)}]{maozmannucci08}
{Maoz}, D., \& {Mannucci}, F. 2008, \mnras, 388, 421

\bibitem[{{Mazzali} {et~al.}(2007){Mazzali}, {R{\"o}pke}, {Benetti}, \&
  {Hillebrandt}}]{mazzali07}
{Mazzali}, P.~A., {R{\"o}pke}, F.~K., {Benetti}, S., \& {Hillebrandt}, W. 2007,
  Science, 315, 825

\bibitem[{{Munari} \& {Renzini}(1992)}]{munarirenzini92}
{Munari}, U., \& {Renzini}, A. 1992, \apjl, 397, L87

\bibitem[{{Nelemans} {et~al.}(2008){Nelemans}, {Voss}, {Roelofs}, \&
  {Bassa}}]{nelemansvoss08}
{Nelemans}, G., {Voss}, R., {Roelofs}, G., \& {Bassa}, C. 2008, \mnras, 388,
  487

\bibitem[{{Nielsen} {et~al.}(2014){Nielsen}, {Gilfanov}, {Bogdan}, {Woods}, \&
  {Nelemans}}]{nielsengilfanov14}
{Nielsen}, M.~T.~B., {Gilfanov}, M., {Bogdan}, A., {Woods}, T.~E., \&
  {Nelemans}, G. 2014, ArXiv e-prints, arXiv:1402.2896

\bibitem[{{Nomoto}(1982)}]{nomoto82}
{Nomoto}, K. 1982, \apj, 253, 798

\bibitem[{{Patat} {et~al.}(2011){Patat}, {Chugai}, {Podsiadlowski}, {Mason},
  {Melo}, \& {Pasquini}}]{patatchugai11}
{Patat}, F., {Chugai}, N.~N., {Podsiadlowski}, P., {et~al.} 2011, \aap, 530,
  A63

\bibitem[{{Patat} {et~al.}(2014){Patat}, {Taubenberger}, {Baade}, {Hoeflich},
  {Maund}, {Reilly}, {Spyromilio}, {Wang}, {Wheeler}, \&
  {Zelaya}}]{patattaubenberger14}
{Patat}, F., {Taubenberger}, S., {Baade}, D., {et~al.} 2014, The Astronomer's
  Telegram, 5830, 1

\bibitem[{{Perlmutter} {et~al.}(1999){Perlmutter}, {Aldering}, {Goldhaber},
  {Knop}, {Nugent}, {Castro}, {Deustua}, {Fabbro}, {Goobar}, {Groom}, {Hook},
  {Kim}, {Kim}, {Lee}, {Nunes}, {Pain}, {Pennypacker}, {Quimby}, {Lidman},
  {Ellis}, {Irwin}, {McMahon}, {Ruiz-Lapuente}, {Walton}, {Schaefer}, {Boyle},
  {Filippenko}, {Matheson}, {Fruchter}, {Panagia}, {Newberg}, {Couch}, \& {The
  Supernova Cosmology Project}}]{perlmutter99}
{Perlmutter}, S., {Aldering}, G., {Goldhaber}, G., {et~al.} 1999, \apj, 517,
  565

\bibitem[{{Perryman} {et~al.}(1997){Perryman}, {Lindegren}, {Kovalevsky},
  {Hoeg}, {Bastian}, {Bernacca}, {Cr{\'e}z{\'e}}, {Donati}, {Grenon},
  {Grewing}, {van Leeuwen}, {van der Marel}, {Mignard}, {Murray}, {Le Poole},
  {Schrijver}, {Turon}, {Arenou}, {Froeschl{\'e}}, \&
  {Petersen}}]{perrymanlindegren97}
{Perryman}, M.~A.~C., {Lindegren}, L., {Kovalevsky}, J., {et~al.} 1997, \aap,
  323, L49

\bibitem[{{Phillips}(1993)}]{ph93}
{Phillips}, M.~M. 1993, \apjl, 413, L105

\bibitem[{{Pickles}(1998)}]{pickles98}
{Pickles}, A.~J. 1998, \pasp, 110, 863

\bibitem[{{Raskin} {et~al.}(2009){Raskin}, {Scannapieco}, {Rhoads}, \& {Della
  Valle}}]{ras09}
{Raskin}, C., {Scannapieco}, E., {Rhoads}, J., \& {Della Valle}, M. 2009, \apj,
  707, 74

\bibitem[{{Riess} {et~al.}(1996){Riess}, {Press}, \& {Kirshner}}]{ri96}
{Riess}, A.~G., {Press}, W.~H., \& {Kirshner}, R.~P. 1996, \apj, 473, 88

\bibitem[{{Riess} {et~al.}(1998){Riess}, {Filippenko}, {Challis},
  {Clocchiatti}, {Diercks}, {Garnavich}, {Gilliland}, {Hogan}, {Jha},
  {Kirshner}, {Leibundgut}, {Phillips}, {Reiss}, {Schmidt}, {Schommer},
  {Smith}, {Spyromilio}, {Stubbs}, {Suntzeff}, \& {Tonry}}]{riess98}
{Riess}, A.~G., {Filippenko}, A.~V., {Challis}, P., {et~al.} 1998, \aj, 116,
  1009

\bibitem[{{Rupen} {et~al.}(2008){Rupen}, {Mioduszewski}, \&
  {Sokoloski}}]{rupenmioduszewski08}
{Rupen}, M.~P., {Mioduszewski}, A.~J., \& {Sokoloski}, J.~L. 2008, \apj, 688,
  559

\bibitem[{{Rushton} {et~al.}(2010){Rushton}, {Kaminsky}, {Lynch}, {Pavlenko},
  {Evans}, {Eyres}, {Woodward}, {Russell}, {Rudy}, {Sitko}, \&
  {Kerr}}]{rushtonkaminsky10}
{Rushton}, M.~T., {Kaminsky}, B., {Lynch}, D.~K., {et~al.} 2010, \mnras, 401,
  99

\bibitem[{{Schaefer}(2010)}]{schaefer10}
{Schaefer}, B.~E. 2010, \apjs, 187, 275

\bibitem[{{Schlafly} \& {Finkbeiner}(2011)}]{schlaflyfinkbeinerSFD11}
{Schlafly}, E.~F., \& {Finkbeiner}, D.~P. 2011, \apj, 737, 103

\bibitem[{{Shappee} \& {Stanek}(2011)}]{shappeestanek11}
{Shappee}, B.~J., \& {Stanek}, K.~Z. 2011, \apj, 733, 124

\bibitem[{{Shen} \& {Bildsten}(2014)}]{shenbildsten14}
{Shen}, K.~J., \& {Bildsten}, L. 2014, \apj, 785, 61

\bibitem[{{Shen} {et~al.}(2012){Shen}, {Bildsten}, {Kasen}, \&
  {Quataert}}]{shenbildsten12}
{Shen}, K.~J., {Bildsten}, L., {Kasen}, D., \& {Quataert}, E. 2012, \apj, 748,
  35

\bibitem[{{Sherrington} \& {Jameson}(1983)}]{sherringtonjameson83}
{Sherrington}, M.~R., \& {Jameson}, R.~F. 1983, \mnras, 205, 265

\bibitem[{{Skopal}(2014)}]{skopal14}
{Skopal}, A. 2014, ArXiv e-prints, arXiv:1402.6126

\bibitem[{{Skrutskie} {et~al.}(2006){Skrutskie}, {Cutri}, {Stiening},
  {Weinberg}, {Schneider}, {Carpenter}, {Beichman}, {Capps}, {Chester},
  {Elias}, {Huchra}, {Liebert}, {Lonsdale}, {Monet}, {Price}, {Seitzer},
  {Jarrett}, {Kirkpatrick}, {Gizis}, {Howard}, {Evans}, {Fowler}, {Fullmer},
  {Hurt}, {Light}, {Kopan}, {Marsh}, {McCallon}, {Tam}, {Van Dyk}, \&
  {Wheelock}}]{skru06}
{Skrutskie}, M.~F., {Cutri}, R.~M., {Stiening}, R., {et~al.} 2006, \aj, 131,
  1163

\bibitem[{{Snijders}(1987)}]{snijders87}
{Snijders}, M.~A.~J. 1987, \apss, 130, 243

\bibitem[{{Sokoloski} {et~al.}(2006){Sokoloski}, {Luna}, {Mukai}, \&
  {Kenyon}}]{sokoloskiluna06}
{Sokoloski}, J.~L., {Luna}, G.~J.~M., {Mukai}, K., \& {Kenyon}, S.~J. 2006,
  \nat, 442, 276

\bibitem[{{Straizys} \& {Kuriliene}(1981)}]{straizyskuriliene81}
{Straizys}, V., \& {Kuriliene}, G. 1981, \apss, 80, 353

\bibitem[{{Sullivan} {et~al.}(2010){Sullivan}, {Conley}, {Howell}, {Neill},
  {Astier}, {Balland}, {Basa}, {Carlberg}, {Fouchez}, {Guy}, {Hardin}, {Hook},
  {Pain}, {Palanque-Delabrouille}, {Perrett}, {Pritchet}, {Regnault}, {Rich},
  {Ruhlmann-Kleider}, {Baumont}, {Hsiao}, {Kronborg}, {Lidman}, {Perlmutter},
  \& {Walker}}]{sullivan10}
{Sullivan}, M., {Conley}, A., {Howell}, D.~A., {et~al.} 2010, \mnras, 406, 782

\bibitem[{{Swings} \& {Allen}(1972)}]{swingsallen72}
{Swings}, J.~P., \& {Allen}, D.~A. 1972, \pasp, 84, 523

\bibitem[{{Szkody}(1977)}]{szkody77}
{Szkody}, P. 1977, \apj, 217, 140

\bibitem[{{Tendulkar} {et~al.}(2014){Tendulkar}, {Liu}, {Dupuy}, \&
  {Ca}}]{tendulkarliu14}
{Tendulkar}, S.~P., {Liu}, M.~C., {Dupuy}, T.~J., \& {Ca}, Y. 2014, The
  Astronomer's Telegram, 5789, 1

\bibitem[{{Thoroughgood} {et~al.}(2001){Thoroughgood}, {Dhillon}, {Littlefair},
  {Marsh}, \& {Smith}}]{thoroughgooddhillon01}
{Thoroughgood}, T.~D., {Dhillon}, V.~S., {Littlefair}, S.~P., {Marsh}, T.~R.,
  \& {Smith}, D.~A. 2001, \mnras, 327, 1323

\bibitem[{{Torres}(2010)}]{torres10}
{Torres}, G. 2010, \aj, 140, 1158

\bibitem[{{van den Heuvel} {et~al.}(1992){van den Heuvel}, {Bhattacharya},
  {Nomoto}, \& {Rappaport}}]{vandenheuvelbhattacharya92}
{van den Heuvel}, E.~P.~J., {Bhattacharya}, D., {Nomoto}, K., \& {Rappaport},
  S.~A. 1992, \aap, 262, 97

\bibitem[{{Wang} {et~al.}(2009){Wang}, {Meng}, {Chen}, \& {Han}}]{wangmeng09}
{Wang}, B., {Meng}, X., {Chen}, X., \& {Han}, Z. 2009, \mnras, 395, 847

\bibitem[{{Webbink}(1984)}]{web84}
{Webbink}, R.~F. 1984, \apj, 277, 355

\bibitem[{{Wheeler}(2012)}]{wheeler12}
{Wheeler}, J.~C. 2012, \apj, 758, 123

\bibitem[{{Whelan} \& {Iben}(1973)}]{whe73}
{Whelan}, J., \& {Iben}, I.~J. 1973, \apj, 186, 1007

\bibitem[{{Wizinowich} {et~al.}(2006){Wizinowich}, {Le Mignant}, {Bouchez},
  {Campbell}, {Chin}, {Contos}, {van Dam}, {Hartman}, {Johansson}, {Lafon},
  {Lewis}, {Stomski}, {Summers}, {Brown}, {Danforth}, {Max}, \&
  {Pennington}}]{wizinowichlemignant06}
{Wizinowich}, P.~L., {Le Mignant}, D., {Bouchez}, A.~H., {et~al.} 2006, \pasp,
  118, 297

\bibitem[{{Woudt} {et~al.}(2009){Woudt}, {Steeghs}, {Karovska}, {Warner},
  {Groot}, {Nelemans}, {Roelofs}, {Marsh}, {Nagayama}, {Smits}, \&
  {O'Brien}}]{woudtsteeghs09}
{Woudt}, P.~A., {Steeghs}, D., {Karovska}, M., {et~al.} 2009, \apj, 706, 738

\bibitem[{{Yoon} \& {Langer}(2003)}]{yoonlanger03}
{Yoon}, S.-C., \& {Langer}, N. 2003, \aap, 412, L53

\bibitem[{{Zamanov} {et~al.}(2010){Zamanov}, {Boeva}, {Bachev}, {Bode},
  {Dimitrov}, {Stoyanov}, {Gomboc}, {Tsvetkova}, {Slavcheva-Mihova}, {Spasov},
  {Koleva}, \& {Mihov}}]{zamanovboeva10}
{Zamanov}, R.~K., {Boeva}, S., {Bachev}, R., {et~al.} 2010, \mnras, 404, 381

\bibitem[{{Zheng} {et~al.}(2014){Zheng}, {Shivvers}, {Filippenko}, {Itagaki},
  {Clubb}, {Fox}, {Graham}, {Kelly}, \& {Mauerhan}}]{zhengshivvers14}
{Zheng}, W., {Shivvers}, I., {Filippenko}, A.~V., {et~al.} 2014, \apjl, 783,
  L24

\end{thebibliography}
\end{document}